\documentstyle[aps,prb]{revtex}

\begin{document}
\draft

\twocolumn[\hsize\textwidth\columnwidth\hsize\csname @twocolumnfalse\endcsname

\title{Resonant absorption at the vortex-core states in $d$-wave superconductors }
\author{N. B. Kopnin}
\address{L.D. Landau Institute for Theoretical Physics,
117334 Moscow, Russia;\\
Low temperature Laboratory, Helsinki University of Technology\\
P.O. Box 2200, FIN-02015 HUT, Finland}
\date{\today}
\maketitle

\begin{abstract}
We predict a resonant microwave absorption on collective vortex modes
in a superclean $d$-wave superconductor at low temperatures. 
Energies of the collective modes are multiples
of the distance between the exact quantum levels of bound states in the
vortex core at lower temperatures and involve delocalized states for higher
temperatures. The characteristic resonant frequency is larger than the
cyclotron frequency $\omega _{c}$ but lower than the
Caroli-deGennes-Matricon minigap $\Delta ^{2}/E_{F}$; it has a $\sqrt{H}$
dependence on the magnetic field and decreases down to $\omega _{c}$
with increasing temperature. We calculate the vortex mass as a response to 
a slow acceleration. This mass is equal, by the order of magnitude, to the 
mass of electrons inside the vortex ``core'' with dimensions $\xi $ by 
$\xi \sqrt{H_{c2}/H}$; it increases with temperature. We discuss the universal
flux-flow regime predicted in \cite{KopVol} and show that it exists in a
broader temperature range than it has been originally found.
\end{abstract}
\


\

] \narrowtext

\section{Introduction}

The vortex dynamics in clean superconductors is determined by localized
states in vortex cores. Motion of a vortex excites transitions between the
states; a competition between the relaxation rate $1/\tau $ and the
interlevel spacing controls the proportion between dissipative and reactive
forces experienced by the vortex \cite{KopKr,KL}. In $d$-wave
superconductors, the vortex dynamics is expected to be more intricate due to
a peculiar structure of the vortex-core states. The presence of gap nodes
introduces the most important difference in the structure of core states
compared to an $s$-wave superconductor. As was shown in Refs. \cite
{KopVol,Volovik}, instead of a well-defined quasiclassical
Caroli-deGennes-Matricon \cite{deG} interlevel spacing $\omega _{0}$, the
true quantum states in $d$-wave superconductors have a much smaller
separation between quantum levels, $E_{0}$, which depends on the magnetic
field. As a result, there appears another parameter $E_{0}\tau $ which
influences the vortex dynamics. According to Ref. \cite{KopVol}, a new
universal regime can be reached in superclean superconductors with
longitudinal and transverse components of the conductivity tensor
independent of the relaxation time. It is realized when the relaxation rate
is smaller than the average distance between the quasiclassical energy
levels $\left\langle \omega _{0}\right\rangle \sim T_{c}^{2}/E_{F}$ but
larger than the separation between the true quantum-mechanical states $E_{0}$%
.

In the present paper, we study both steady and oscillatory motion of
vortices at low temperatures $T\ll T_{c}$ and low magnetic fields $H\ll
H_{c2}$ in more detail using the microscopic theory outlined in Ref. \cite
{KopVol}. For a steady flux flow, we find that the universal regime is
realized in a considerably broader region of temperatures and magnetic
fields than what was originally predicted in Ref. \cite{KopVol}. This
conclusion agrees with the results of Ref. \cite{Makhlin}. The relevant
parameter which governs the vortex dynamics is shown to be the energy of
``collective modes'' $E_{{\rm col}}$. We calculate $E_{{\rm col}}$ and show
that it coincides with the true quantum mechanical interlevel spacing $E_{0}$
if the energy of excitations is $\epsilon \ll T_{c}\sqrt{H/H_{c2}}$, and
involves delocalized states for excitations with higher energies $\epsilon
\gg T_{c}\sqrt{H/H_{c2}}$. For an oscillatory motion of vortices which can
be excited by a microwave irradiation, we predict a finite resonant
absorption even for infinitely long relaxation time. The series of
resonances occurs at multiples of $E_{{\rm col}}$ with the resonant
frequencies considerably lower than the average quasiclassical interlevel
distance $T_{c}^{2}/E_{F}$ but higher than the cyclotron frequency $\omega
_{c}$. They have a $\sqrt{H}$ magnetic field dependence at very low
temperatures. For higher temperatures, the series of resonances compresses
down to an absorption edge which approaches $\omega _{c}$ for $T\sim T_{c}$.

\section{Nonequilibrium excitations}

\subsection{Spectrum}

\label{excitations}

We consider superconductors which have the coherence length $\xi $ much
longer than the inverse Fermi momentum, $p_{F}\xi \gg 1$. For these
superconductors, a quasiclassical approximation is appropriate. A
quasiclassical particle moves along a straight line parallel to its momentum
which is conserved with the accuracy $(p_{F}\xi )^{-1}$. The angular
momentum is thus another approximately conserved quantity even if the order
parameter does not have a cylindrical symmetry; it is now a continuous
variable not directly related with labelling the quantum states in the
vortex core. A quasiparticle passing at some distance near a vortex can
become classically localized in the vortex core region. It will have an
energy characterized by the momentum along the vortex axis $p_{z}$, the
momentum direction in the plane perpendicular to the vortex axis, and by the
impact parameter $b=-\mu /p_{\perp }$ coupled to the angular momentum $\mu $.

For a $d$-wave superconductor without admixture of other components, the
order parameter is 
\begin{equation}
\Delta _{{\bf p}}=\Delta _{0{\bf p}}e^{i\phi }=\Delta _{0}\sin (2\alpha
)e^{i\phi }  \label{order/param}
\end{equation}
where $\alpha $ is the angle between ${\bf p}_{\perp }$ and the $x$ axis
which is taken along one of the gap nodes. The modulus of the order
parameter $\Delta _{0}=\Delta _{0}(\rho ,\phi );$ at large distances, $%
\Delta _{0}=\Delta _{\infty }$. We take the $z$ axis along the vortex
circulation, ${\bf z}={\rm sign}\,(e){\bf H}/H$, and denote by ${\bf v}%
_{\perp }$ the quasiparticle velocity in the $(x,y)$ plane. If $\rho $ and $%
\phi $ are the distance and the azimuthal angle in the cylindrical frame,
the impact parameter $b$ of a quasiparticle moving through the point $(\rho
,\phi )$ and the distance $s$ along the trajectory are 
\begin{equation}
b=\rho \sin (\phi -\alpha );s=\rho \cos (\phi -\alpha ).
\end{equation}

The quasiparticle energy can be found using the quasiclassical Green
function technique in the same way as it has been done for $s$-wave
superconductors in Ref.\cite{KrPe}. Here we summarize the most important
properties of the energy spectrum of a quasiclassical particle in the vortex
core. It is discussed in more detail in Appendix. The quasiclassical
spectrum has several branches belonging to various values of the radial
quantum number $n$. The anomalous branch with $n=0$ crosses zero of energy
as a function of the impact parameter. For small impact parameters $b\ll \xi
/|\sin (2\alpha )|$, 
\begin{equation}
\epsilon _{0}(\alpha ,b)=\frac{2\Delta _{\infty }^{2}\sin ^{2}(2\alpha )Lb}{%
v_{\perp }}\,+\frac{p_{\perp }b\omega _{c}}{2}  \label{E/small/b}
\end{equation}
where $L=\ln \left[ 1/|\sin (2\alpha )|\right] $ for $b\ll \xi $ and $L=\ln
\left[ \xi v_{\perp }/bv_{F}|\sin (2\alpha )|\right] $ for $\xi \ll b\ll \xi
/|\sin (2\alpha )|$; the second term is the magnetic energy where $\omega
_{c}=|e|H/m_{\perp }c$ is the cyclotron frequency with $m_{\perp }=p_{\perp
}/v_{\perp }$. Modulus of charge appears due to the choice of the $z$ axis.
Note that the magnetic energy has the energy level spacing \cite{Hansen}
equal to the Larmor frequency $\omega _{c}/2$ rather than just to $\omega
_{c}$ as it would be the case for the Landau spectrum in the normal state.
This is because the wave function is localized at distances of the general
order of $\xi $ rather than at the magnetic radius $\rho _{H}\sim p_{\perp
}c/eH$ which would be the case in the normal state. Near the gap nodes,
i.e., when $\sin (2\alpha )$ is small, the low-lying states with energies
much below the gap at infinity, $\epsilon \ll \Delta _{0}|\sin 2\alpha |$,
correspond to $b\ll \xi /|\sin 2\alpha |$. The derivative of the energy with
respect to the angular momentum, $\omega _{0}(\alpha )=p_{\perp
}^{-1}[\partial \epsilon _{0}(\alpha ,b)/\partial b]$, is the analogue of
the Caroli-deGennes-Matricon minigap \cite{deG} in an $s$-wave
superconductor.

For large impact parameters, $b\gg \xi /|\sin 2\alpha |$, the energy is 
\begin{equation}
\epsilon _{0}(b)={\rm sign}\,(b)\left( \Delta _{\infty }|\sin (2\alpha )|-%
\frac{v_{\perp }}{4|b|}\right) +\frac{p_{\perp }b\omega _{c}}{2}.
\label{spec/nodes3}
\end{equation}
The particles are localized on lines passing through the vortex
perpendicular to the momentum direction. The localization length along the
trajectory is $s\sim \sqrt{\xi b/|\sin (2\alpha )|}\ll b$. A similar
spectrum was considered in Ref. \cite{Makhlin}.

Eqs. (\ref{E/small/b}) and (\ref{spec/nodes3}) can be combined into a single
interpolation formula 
\begin{equation}
\epsilon _{0}(b)=\frac{\Delta _{\infty }^{2}\sin ^{2}(2\alpha )b}{|b|\Delta
_{\infty }|\sin (2\alpha )|+\beta v_{\perp }}+\frac{p_{\perp }b\omega _{c}}{2%
}  \label{spec}
\end{equation}
where $\beta $ is a constant. The fit to the large-$b$ region of energy
gives $\beta =1/4$; the best fit for low-$b$ region is reached with $\beta
=1/(2L)$. We define an angle $\alpha _{0}$ at which the particle energy $%
\tilde{\epsilon}=\epsilon _{0}(b)-p_{\perp }b\omega _{c}/2$ is equal to the
gap $\tilde{\epsilon}=\Delta _{\infty }|\sin 2\alpha _{0}|$; the angle $%
\alpha _{0}$ is close to one of the nodes if $\tilde{\epsilon}\ll \Delta
_{\infty }$. It is a quasiparticle moving at an angle $\alpha >\alpha _{0}$
which is classically localized.

Other branches with nonzero radial quantum numbers $n$ are separated from $%
\epsilon _{0}(b)$ by energies of the order of $\Delta _{\infty }|\sin
(2\alpha )|$. The number of branches with energies between $\tilde{\epsilon}$
and $\Delta _{\infty }|\sin (2\alpha )|$ is 
\begin{equation}
\nu \sim \frac{\Delta _{\infty }|\sin (2\alpha )|}{\sqrt{[\Delta _{\infty
}^{2}\sin ^{2}(2\alpha )-\tilde{\epsilon}^{2}]}}.  \label{Number}
\end{equation}

Eqs. (\ref{spec/nodes3}) and (\ref{spec}) are quantitatively correct as long
as $b\ll \rho _{{\rm max}}$ where $\rho _{{\rm max}}\sim \xi \sqrt{H_{c2}/H}$
is the distance between vortices. Such distances are, in principle,
accessible for a particle which moves near the direction of a gap node
within the angle $\alpha \sim \sqrt{H/H_{c2}}$. We shall see, however, that
the condition $b\ll \rho _{{\rm max}}$ is fulfilled for energies $\epsilon
\ll \Delta _{\infty }\sqrt{H/H_{c2}}$.

A classically localized particle is not necessarily localized in a strict
quantum-mechanical sense. This was also found in numerical calculations of
Ref. \cite{Tesanovic}. The true quantum levels can be obtained from the
semiclassical Bohr-Sommerfeld quantization rule \cite
{KopVol,KV/dos,Vol/quant} 
\begin{equation}
\oint \mu (\alpha )\,d\alpha =2\pi \left( m+\frac{1}{2}\right) ;
\label{BS/quant}
\end{equation}
$m$ is an integer, $\frac{1}{2}$ appears because the single-particle wave
function changes its sign after encircling a single-quantum vortex. The
angular momentum $\mu (\alpha )$ is expressed through the quasiparticle
energy according to Eq. (\ref{spec}). Consider first an energy $\epsilon \ll
\Delta _{\infty }\sqrt{H/H_{c2}}$. The integral in Eq. (\ref{BS/quant})
converges according to Eq. (\ref{spec}) and is determined by angles $\alpha
\sim \sqrt{H/H_{c2}}$. For this region of angles, the spectrum is actually
given by the exact equation (\ref{E/small/b}): $\epsilon _{0}(\mu )=-\omega
_{0}(\alpha )\mu $ where 
\begin{equation}
\omega _{0}(\alpha )=8\Omega _{0}\alpha ^{2}+\omega _{c}/2
\label{spec/nodes2}
\end{equation}
with 
\begin{equation}
\Omega _{0}=(\Delta _{\infty }^{2}/v_{\perp }p_{\perp })\ln [(\Delta
_{\infty }/\epsilon )\sqrt{H/H_{c2}}].
\end{equation}
The characteristic impact parameters are of the order of $b\sim \epsilon
/p_{\perp }\omega _{c}$, i.e., $b\ll\rho _{{\rm max}}$. We obtain $%
E_{m}=E_{0}(m+1/2)$ where 
\begin{equation}
E_{0}=\sqrt{\Omega _{0}\omega _{c}}.  \label{E/0}
\end{equation}

A qualitative expression for the spectrum with the $\sqrt{H}$ dependence on 
the magnetic field similar to Eq. (\ref{E/0}) was found in 
Ref. \cite{Volovik} neglecting the magnetic energy. The argumentation was 
that the divergence of $\omega _0^{-1}(\alpha )$ without the magnetic energy 
should be cut off at such angles 
$\alpha$ that the characteristic distance of a qusiparticle from the vortex 
axis is of the order of the intervortex distance $\rho _{{\rm max}}$; 
here the corrections to the quasiparticle energy induced
by neighboring vortices is just of the order of $\omega _0(\alpha )$. We
see, however, that Eq. (\ref{E/0}) is obtained exactly if one takes into
account the magnetic quantization. We find simultaneously that the particles
with energies $\epsilon \ll \Delta \sqrt{H/H_{c2}}$ are acually
localized in the vortex core since $b\ll \rho _{{\rm max}}$; thus the
corrections from neighboring vortices are not important. For larger energies 
$\epsilon \sim \Delta _{\infty }\sqrt{H/H_{c2}}$, the localization radius
becomes of the order of $\rho _{\rm max}$. These particles can not
be considered as localized any more because of the presence of other
vortices. For higher energies $\epsilon \gg \Delta _{\infty }\sqrt{H/H_{c2}}$%
, the part with extended trajectories $\alpha <\alpha _{0}$ in Eq. (\ref
{BS/quant}) dominates, and thus quasiparticles are no longer localized in
the core even for an isolated vortex.

The average quasiclassical energy-level spacing $\left\langle \omega
_{0}\right\rangle \sim \Omega _{0}\sim \Delta _{\infty }^{2}/E_{F}$
determines the parameter $\Omega _{0}\tau $ which controls the vortex
dynamics in clean $s$-wave superconductors with $\tau \gg T_{c}^{-1}$. The
value $\Omega _{0}\tau \sim 1$ separates the so called moderately clean
limit, $\Omega _{0}\tau \ll 1$, with a highly dissipative vortex dynamics
from the superclean limit, $\Omega _{0}\tau \gg 1$, where dissipation is
small.

In $d$-wave superconductors, situation is more intriguing. It was predicted
in Ref. \cite{KopVol} that a large dissipation persists for much longer $%
\ell $. In the present paper we show that, for a steady vortex motion, the
relevant parameter which marks the transition between dissipative and
nondissipative dynamics is $E_{{\rm col}}\tau $ where $E_{{\rm col}}$ is the
temperature dependent characteristic energy of ``collective modes'' induced
by moving vortex. For low temperatures, it coincides with the energy of
``bound states'' determined by Eqs. (\ref{BS/quant}), (\ref{E/0}): $E_{{\rm %
col}}=E_{0}\ll \Omega _{0}$. The energy $E_{{\rm col}}$ decreases down to $%
\omega _{c}$ when the temperature approaches $T_{c}$. The parameter $\omega
_{c}\tau $ is much smaller than $\Omega _{0}\tau $ since $\omega _{c}\sim
(H/H_{c2})\Omega _{0}$. We have thus a hierarchy of energies $\omega _{c}\ll
E_{{\rm col}}\ll \Omega _{0}$. To get the parameter which controls the
vortex dynamics, these energies should be compared either with the
relaxation rate $1/\tau $ for a steady motion of vortices or with ${\rm max}%
\,(1/\tau ,\,\omega )$ where $\omega $ is the characteristic frequency of
vortex oscillations. Large values of $\Omega _{0}\tau $ require a very high
purity of samples: the mean free path should be $\ell \gg (T_{c}/E_{F})\xi $%
. Nevertheless, there are experimental evidences that such a regime can be
realized in practice. \cite{Ong}

\subsection{Balance of forces on a moving vortex}

The force on a vortex from the environment where it moves is the momentum
transferred from the excitations. The exact microscopic expression for the
force has been derived in Ref. \cite{KL}. It can equivalently be presented
as a force produced by quasiclassical particles with a distribution $f$,
characterized by canonically conjugated coordinate $\alpha $ and angular
momentum $\mu $ \cite{Stone}. The contribution from classically localized
states is 
\begin{equation}
{\bf F}_{{\rm env}}^{({\rm loc})}=-\sum_{n}\int \frac{dp_{z}}{2\pi }\,\frac{%
d\mu \,d\alpha }{2\pi }\frac{\partial {\bf p}_{n}}{\partial t}f.
\end{equation}
Using $\partial {\bf p}/\partial t=-\nabla \epsilon _{n}=[\hat{{\bf z}}%
\times {\bf p}](\partial \epsilon _{n}/\partial \mu )$ we get 
\begin{eqnarray}
{\bf F}_{{\rm env}}^{({\rm loc})} &=&-\sum_{n}\int \frac{dp_{z}}{2\pi }\frac{%
d\mu \,d\alpha }{2\pi }[\hat{{\bf z}}\times {\bf p}_{\perp }]\frac{\partial
\epsilon _{n}}{\partial \mu }\delta f_{n}  \nonumber \\
&=&\pi N[\hat{{\bf z}}\times {\bf u}]\left\langle \sum_{n}\int \gamma _{H}%
\frac{df_{0}}{d\epsilon }\,\frac{\partial \epsilon _{n}}{\partial \mu }%
\,d\mu \right\rangle _{F}  \nonumber \\
&&-\pi N{\bf u}\left\langle \sum_{n}\int \gamma _{O}\frac{df_{0}}{d\epsilon }%
\frac{\partial \epsilon _{n}}{\partial \mu }\,d\mu \right\rangle _{F}.
\label{F/env}
\end{eqnarray}
We take the nonequilibrium distribution function in the form 
\begin{equation}
\delta f=-\frac{df_{0}}{d\epsilon }\left( ({\bf up}_{\perp }{\bf )}\gamma
_{H}+\left( \left[ {\bf u\times p}_{\perp }\right] {\bf z}\right) \gamma
_{O}\right)  \label{delta/f}
\end{equation}
where $f_{0}$ is the equilibrium Fermi distribution, the factors $\gamma
_{O} $ and $\gamma _{H}$ describe the longitudinal and transverse responses
to the vortex velocity ${\bf u}$, and $N$ is the electron density; $%
\left\langle \cdots \right\rangle _{F}$ is the average over the whole Fermi
surface which includes averaging over $\alpha $. We assume here a Fermi
surface with not less than the tetragonal symmetry in the plane
perpendicular to the vortex axis. For simplicity, we consider only the
particle-like Fermi surface, the generalization for a surface with both
particle-like and hole-like parts is given later.

The force from classically delocalized states is \cite{KL} 
\begin{eqnarray}
{\bf F}_{{\rm env}}^{({\rm del})}&=&\pi N [\hat{{\bf z}}\times {\bf u}%
]\left\langle \int_{{\rm del}}\gamma _{H}\frac{df_{0}}{d\epsilon }%
\,d\epsilon \right\rangle _{F}  \nonumber \\
&&-\pi N{\bf u}\left\langle \int_{{\rm del}}\gamma _{O}\frac{df_{0}}{%
d\epsilon }\,d\epsilon \right\rangle _{F} .  \label{F/env/del}
\end{eqnarray}

The force ${\bf F}_{{\rm env}}$ should be balanced by the Lorentz force $%
{\bf F}_{L}={\rm sign}\,(e)\phi _{0}[{\bf j}\times \hat{{\bf z}}]/c$ where $%
\phi _{0}=\pi c/|e|$ is the flux quantum. Since the average electric field
is ${\bf E}=[{\bf B}\times {\bf u}]/c$, the force balance ${\bf F}_{L}+{\bf F%
}_{{\rm env}}=0$ gives a linear relation between the transport current and
the electric field. The proportionality coefficients are the Ohmic and Hall
conductivities. The Ohmic conductivity is 
\begin{eqnarray}
\sigma _{O} &=&\frac{N|e|c}{B}\left\langle \sum_{n}\int \gamma _{O}\frac{%
df_{0}}{d\epsilon }\,\frac{\partial \epsilon _{n}}{\partial \mu }\,d\mu
\right\rangle _{F}  \nonumber \\
&&+\frac{N|e|c}{B}\left\langle \int_{{\rm del}}\gamma _{O}\frac{df_{0}}{%
d\epsilon }\,d\epsilon \right\rangle _{F}.  \label{sigmaO1}
\end{eqnarray}
For the Hall conductivity $\sigma _{H}$ one has the same expression with $%
\gamma _{H}$ instead of $\gamma _{O}$ and $|e|$ replaced with $e$.

In Galilean invariant systems ${\bf j}=N_{s}e{\bf v}_{s}$, and the force
balance is sometimes presented in terms of the Magnus force from the
superfluid component and the friction plus transverse forces from the normal
component 
\[
\pi N_{s}[({\bf v}_{s}-{\bf u})\times \hat{{\bf z}}]-D{\bf u}-D^{\prime }[%
\hat{{\bf z}}\times {\bf u}]=0.
\]
In such representation, the Magnus force includes the Lorentz force and a
part of the transverse force ${\bf F}_{{\rm env},\perp }$. The constants $D$
and $D^{\prime }$ are expressed through the conductivities as 
\[
\sigma _{O}=\frac{c^{2}D}{\phi _{0}B};~\sigma _{H}={\rm sign}\,(e)\frac{%
c^{2}(\pi N_{s}-D^{\prime })}{\phi _{0}B}.
\]

\section{Distribution function}

\subsection{Kinetic equation}

The kinetic equation for the distribution function $f(t,\alpha ,\mu )$ for a
system of fermions characterized by canonically conjugated variables $\mu $
and $\alpha $ has the form \cite{Stone} 
\begin{equation}
{\frac{\partial f}{\partial t}}+{\frac{\partial f}{\partial \alpha }}{\frac{%
\partial \epsilon _{n}}{\partial \mu }}-{\frac{\partial \epsilon _{n}}{%
\partial \alpha }}{\frac{\partial f}{\partial \mu }}=-{\frac{f-f_{0}}{\tau
_{n}}}.  \label{KE}
\end{equation}
A particle is classically localized if it has an energy $\tilde{\epsilon}%
<\Delta _{\infty }$ and moves at an angle $\alpha >\alpha _{0}$ counted from
one of the nodes. For localized quasiparticles, the derivative $\partial
\epsilon _{n}/\partial \mu =-\omega _{n}(\alpha )$ is the interlevel
spacing. The collision integral is written in a $\tau $-approximation. This
equation can also be derived microscopically. \cite{KL}

If the vortex moves with a velocity ${\bf u}$ with respect to the heat bath,
the Doppler shift of the energy $\epsilon \rightarrow \epsilon (\mu ,\alpha
)-{\bf p}{\bf u}$ produces the ``driving force'' $(\partial \epsilon
_{n}/\partial \alpha )=[{\bf u}\times {\bf p}_{\perp }]\hat{{\bf z}}$ acting
on quasiparticles. The third term in the l.h.s. of kinetic equation (\ref{KE}%
) contains this force multiplied by $\partial f_{0}/\partial \mu =-\omega
_{n}(\alpha )(df_{0}/d\epsilon )$. We look for a solution in the form $%
f-f_{0}=\delta f$ where $\delta f=\delta f(\epsilon ,\alpha )$ is
independent of $b$. Eq. (\ref{KE}) gives 
\begin{equation}
\frac{\partial \delta f}{\partial \alpha }-([{\bf u}\times {\bf p}_{\perp }]%
\hat{{\bf z}})\frac{df_{0}}{d\epsilon }=U(\alpha )\delta f  \label{KE1}
\end{equation}
where 
\begin{equation}
U_{n}(\alpha )=\left( -i\omega +\frac{1}{\tau }\right) \omega
_{n}^{-1}(\alpha ).  \label{U/loc}
\end{equation}
Here we assume that the vortex velocity has a form ${\bf u}={\bf u}_{\omega
}e^{-i\omega t}$ where $\omega \ll \Delta $.

Introducing the longitudinal and transverse responses to the vortex velocity
according to Eq. (\ref{delta/f}) and taking into account that, for the
tetragonal symmetry, the responses do not depend on the direction of the
vortex motion with respect to the crystal lattice, one finds two coupled
first-order differential equations for $\gamma _{O}(\alpha )$ and $\gamma
_{H}(\alpha )$: 
\begin{eqnarray}
\frac{\partial \gamma _{O}}{\partial \alpha }-\gamma _{H}-U(\alpha )\gamma
_{O}+1 &=&0,  \nonumber \\
\frac{\partial \gamma _{H}}{\partial \alpha }+\gamma _{O}-U(\alpha )\gamma
_{H} &=&0.  \label{eq-gamma}
\end{eqnarray}

A quasiparticle is delocalized either for energies $\tilde{\epsilon}>\Delta
_{\infty }$ or for angles $\alpha <\alpha _{0}$. For a homogeneous magnetic
field the distribution function of delocalized electrons was shown \cite{KL}
to satisfy Eq. (\ref{KE1}) where $\omega _{n}$ is replaced with $g\omega _{c}
$. Here $g$ is the number of states for a particle with given $\tilde{%
\epsilon}$, $\alpha $, and $b$ at large distances from the vortex, i.e., the
quasiclassical Green function $g$: 
\[
g=\tilde{\epsilon}/\sqrt{\tilde{\epsilon}^{2}-\Delta _{0}^{2}\sin
^{2}2\alpha }.
\]
The distribution function thus has the form of Eq. (\ref{delta/f}) with $%
\gamma _{O}$ and $\gamma _{H}$ satisfying Eqs. (\ref{eq-gamma}) where now 
\begin{equation}
U(\alpha )=\left( -i\omega +\frac{1}{\tau }\right) \left( \omega
_{c}g\right) ^{-1}.  \label{U/del}
\end{equation}
Since $\gamma _{O}$ and $\gamma _{H}$ obey first-order differential
equations they are continuous functions at $\alpha =\alpha _{0}$. For $%
\tilde{\epsilon}>\Delta _{\infty }$, the potential is given by Eq. (\ref
{U/del}) in the whole region of angles.

\subsection{Static response}

Equations (\ref{eq-gamma}) can be easily solved. We have $\gamma _{H}(\alpha
)={\rm Re}\,W$; $\gamma _{O}(\alpha )={\rm Im}\,W$ where 
\[
W=e^{\left[ i\alpha +F\left( \alpha \right) \right] }\left(
C-i\int_{0}^{\alpha }e^{-\left[ i\alpha ^{\prime }+F\left( \alpha ^{\prime
}\right) \right] }\,d\alpha ^{\prime }\right) 
\]
with 
\[
F(\alpha )=\int_{0}^{\alpha }U(\alpha ^{\prime })\,d\alpha ^{\prime }. 
\]
In the moderately clean limit, $\Omega _{0}\tau \ll 1$, the potential $%
U(\alpha )$ is always large, and we obtain the local solution as in an $s$%
-wave superconductor \cite{KL} 
\begin{equation}
\gamma _{O}(\alpha )=\omega _{0}(\alpha )\tau ,~\gamma _{H}(\alpha )=[\omega
_{0}(\alpha )\tau ]^{2}.  \label{gammas/mod}
\end{equation}

In the superclean limit $\Omega _{0}\tau \ll 1$, the potential $U(\alpha )$
is small almost everywhere except for vicinities of the gap nodes where it
becomes large. Consider first energies $\epsilon \ll \Delta _{\infty }\sqrt{%
H/H_{c2}}$ and find the distribution function for the anomalous branch $n=0$
in the region of angles $\alpha $ not specifically close to the gap nodes.
It is this branch which is only excited at temperatures $T\ll T_{c}\sqrt{%
H/H_{c2}}$. The overall behavior of the distribution function is to the
highest extent determined by what happens in a close vicinity of the gap
nodes. It is this region which is responsible for the whole build-up of
nonequilibrium distribution of excitations.

For angles larger than $\alpha \sim \sqrt{H/H_{c2}}$ from the nodes one can
neglect the potential $U$. As a result, one has for $\delta \alpha <\alpha
<\pi /2-$ $\delta \alpha $ where $\sqrt{H/H_{c2}}\ll \delta \alpha \ll 1$, 
\begin{eqnarray}
\gamma _{O}(\alpha ) &=&A\cos \alpha +B\sin \alpha ;  \nonumber \\
\;\gamma _{H}(\alpha ) &=&1-A\sin \alpha +B\cos \alpha .
\label{gamma/solution}
\end{eqnarray}
The constants $A$ and $B$ can be found by matching with the solution in the
vicinity of the node, $\alpha \ll 1$, where $\gamma _{O,H}(\alpha )\propto
e^{F(\alpha )}$. This provides the boundary condition $\gamma _{O,H}(+\delta
\alpha )=e^{2\lambda }\gamma _{O,H}(-\delta \alpha )$ across the node at $%
\alpha =0$. Here $2\lambda =F(+\delta \alpha )-F(-\delta \alpha )$. For such
energies, the region of angles $\alpha <\alpha _{0}$ with delocalized
trajectories is not important. The integral for $\lambda $ converges and is
determined by angles $\alpha \sim \sqrt{H/H_{c2}}$. This range of angles
sets the width of the transition region near a gap node where the
distribution function jumps from its value at $\alpha =-\delta \alpha $ to
its value at $\alpha =+\delta \alpha $. We obtain using Eq. (\ref
{spec/nodes2}) 
\begin{equation}
\lambda =\tau ^{-1}\int_{0}^{\infty }\frac{d\alpha }{\omega _{0}(\alpha )}=%
\frac{\pi }{4E_{0}\tau }.
\end{equation}

The solution for $\gamma _{O,H}$ should be periodically continued to the
rest of angles with the period $\pi /2$ since the response function has the
same tetragonal symmetry as the underlying system: $\gamma (\pi /2-\delta
\alpha )=\gamma (-\delta \alpha )$. Together with the above boundary
condition, this gives 
\begin{equation}
A=\frac{e^{\lambda }\sinh \lambda }{2\sinh ^{2}\lambda +1};\;B=\frac{%
e^{-\lambda }\sinh \lambda }{2\sinh ^{2}\lambda +1}.  \label{A,B}
\end{equation}
We have after averaging over the azimuthal angle $\alpha $ 
\begin{equation}
\left\langle \gamma _{H}\right\rangle =1-\frac{4}{\pi }\frac{\tanh
^{2}\lambda }{1+\tanh ^{2}\lambda },~\left\langle \gamma _{O}\right\rangle =%
\frac{4}{\pi }\frac{\tanh \lambda }{1+\tanh ^{2}\lambda }.  \label{gammas}
\end{equation}

For energies $\epsilon \gg \Delta _{\infty }\sqrt{H/H_{c2}}$, the region of
angles $\alpha <\alpha _{0}$ with extended trajectories dominates; now it is 
$\alpha _{0}$ which determines the width of the transition region where $%
\gamma _{O}$ and $\gamma _{H}$ jump. Compact expressions can be obtained if
we assume that $U(\alpha )$ is independent of $\alpha $ for $\alpha <\alpha
_{0}$. For simplicity, we replace $g\omega _{c}\tau $ with $\omega _{c}\tau $%
. This does not change the results qualitatively. We have for $\alpha
<\alpha _{0}$%
\begin{eqnarray}
\gamma _{O} &=&d_{1}+e^{\alpha /\omega _{c}\tau }\left( C_{1}\sin \alpha
+C_{2}\cos \alpha \right) ,  \nonumber \\
\gamma _{H} &=&d_{2}+e^{\alpha /\omega _{c}\tau }\left( C_{1}\cos \alpha
-C_{2}\sin \alpha \right)   \label{gammas/del}
\end{eqnarray}
where 
\[
d_{1}=\frac{\omega _{c}\tau }{\omega _{c}^{2}\tau ^{2}+1};\;d_{2}=\frac{%
\omega _{c}^{2}\tau ^{2}}{\omega _{c}^{2}\tau ^{2}+1}.
\]
We can use Eq. (\ref{gamma/solution}) for angles $\alpha >\alpha _{0}+\delta
\alpha $. Since the contribution to $\lambda $ form the region of angles $%
\alpha \sim \delta \alpha $ is much smaller than that from the angles $%
\alpha \sim \alpha _{0}$, we can extend Eq. (\ref{gamma/solution}) to the
angles $\alpha =\alpha _{0}$ and $\alpha =\pi /2-\alpha _{0}$. First, we
match equations (\ref{gamma/solution}) and (\ref{gammas/del}) at $\alpha
=\alpha _{0}$. Another condition is obtained by matching Eqs. (\ref
{gammas/del}) taken at $\alpha =-\alpha _{0}$ with Eq. (\ref{gamma/solution}%
) taken at $\alpha =\pi /2-\alpha _{0}$. We obtain four equations for four
constants $A$, $B$, $C_{1}$, and $C_{2}$ where now $\lambda =\alpha
_{0}/\omega _{c}\tau =\epsilon /(2\Delta _{\infty }\omega _{c}\tau )$.

If $\lambda $ is not considerably smaller than unity, we can neglect 
$\alpha _{0}$ in the boundary conditions, and obtain 
\begin{eqnarray*}
A &=&\left( 1-d_{2}\right) \frac{e^{\lambda }\sinh \lambda }{2\sinh
^{2}\lambda +1}-d_{1}\frac{e^{-\lambda }\sinh \lambda }{2\sinh ^{2}\lambda +1%
}; \\
B &=&\left( 1-d_{2}\right) \frac{e^{-\lambda }\sinh \lambda }{2\sinh
^{2}\lambda +1}+d_{1}\frac{e^{\lambda }\sinh \lambda }{2\sinh ^{2}\lambda +1}%
.
\end{eqnarray*}
For values of $\lambda \sim 1$ and larger, one automatically has $\omega
_{c}\tau \ll 1$. As a result, both $d_{1}$ and $d_{2}$ are small, and we
recover Eqs. (\ref{A,B}) and (\ref{gammas}) with the new definition for $%
\lambda $. The coefficients $C_1$ and $C_2$ are in this case
\[ C_1= \frac{\cosh \lambda}{2\sinh ^2\lambda +1};~
C_2= \frac{\sinh \lambda}{2\sinh ^2\lambda +1}.
\]
We see that the distribution function for energies 
$\Delta _\infty \sqrt{H/H_{c2}}\ll \epsilon \ll \Delta _\infty$ is not small 
even for delocalized trajectories $\alpha <\alpha _0$.

The contribution from the region of angles $\alpha \sim 1$ to Eq. (\ref
{sigmaO1}) and to the corresponding equation for $\sigma _{H}$ decreases
together with $\lambda $ as the parameter $\omega _{c}\tau $ increases. For $%
\lambda \ll 1$, the contribution from angles $\alpha \sim \alpha _{0}$ where
the gap $\Delta _{\infty }|\sin (2\alpha )|$ is of the order of temperature
becomes important. In this case, we obtain from Eqs.(\ref{gamma/solution}) 
and (\ref{gammas/del})   
\begin{eqnarray*}
A=\lambda (1-d_2-d_1)+\alpha _0 (1-d_2+d_1);\\
B=\lambda (1-d_2+d_1)-\alpha _0 (1-d_2-d_1);\\
C_1= 1-d_2+\lambda d_1;~ C_2=\lambda (1-d_2)-d_1.
\end{eqnarray*}
As a result, 
\begin{equation}
\left\langle \gamma _{O}\right\rangle =\frac{4}{\pi }\lambda ;\;\left\langle
\gamma _{H}\right\rangle =1.  \label{gammas/extreme}
\end{equation}
We can thus use Eq. (\ref{gammas}) within the whole range of $\lambda $ and
combine the two results for $\lambda $ in different regions of energy into a
single interpolation expression 
\begin{equation}
\lambda =\frac{\pi }{4E_{{\rm col}}\tau }  \label{lambda1}
\end{equation}
where $E_{{\rm col}}$ has the meaning of a characteristic energy of
``collective modes'' 
\begin{equation}
E_{{\rm col}}=\left[ \frac{1}{\sqrt{\Omega _{0}\omega _{c}}}+\frac{%
2|\epsilon |}{\pi \Delta _{\infty }\omega _{c}}\right] ^{-1}
\label{exact/levels}
\end{equation}
which we discuss later.

For energies $\epsilon >\Delta _{\infty }$ we have from Eqs. (\ref
{gammas/del}) $C_{1}=C_{2}=0$ and 
\begin{equation}
\gamma _{O}=d_{1};\gamma _{H}=d_{2}.
\end{equation}

\subsection{Frequency-dependent response}

Assume that the external frequency $\omega \ll \Omega _{0}$. In this limit,
the solution of the kinetic equation (\ref{KE}) has the form of Eqs. (\ref
{gammas},\ref{lambda1}) where $1/\tau $ in Eq. (\ref{lambda1}) is replaced
with $1/\tau -i\omega $. Consider first the case $\omega \gg \omega _{c}$.
The factors $d_{1}\sim \omega _{c}/\omega $ and $d_{2}\sim \omega
_{c}^{2}/\omega ^{2}$ are small. In the limit $\omega \tau \to \infty $, the
longitudinal response is 
\begin{equation}
\left\langle \gamma _{O}\right\rangle =i\frac{4}{\pi }\frac{\tan \lambda
^{\prime }}{\tan ^{2}\lambda ^{\prime }-1}
\end{equation}
where 
\begin{equation}
\lambda ^{\prime }=\frac{\pi \omega }{4E_{{\rm col}}}.  \label{lambda/prime}
\end{equation}
The response has poles at $\lambda ^{\prime }=\pi /4+\pi k/2$ or at
frequencies $\omega =\omega _{k}$ where $\omega _{k}=E_{{\rm col}}(1+2k)$
and $E_{{\rm col}}$ is given by Eq. (\ref{exact/levels}). Harmonics with $%
k\ne 0$ appear due to the absence of the axial symmetry. Note that $E_{{\rm %
col}}\gg \omega _{c}$. For low temperatures $T/T_{c}\ll \sqrt{H/H_{c2}}$,
the eigen frequencies are independent of temperature: $\omega
_{k}=E_{0}(1+2k)$. These poles are the collective modes of electrons
involved into the vortex motion. For low energies $\epsilon \ll \Delta
_{\infty }\sqrt{H/H_{c2}}$, these modes coincide with multiples of the
distance between the true quantum mechanical energy levels determined by Eq.
(\ref{BS/quant}). For higher energies, delocalized quasiparticles are
involved into the vortex motion.

\section{Conductivities}

\subsection{Steady motion}

Since $\gamma _{O}$ and $\gamma _{H}$ do not depend on $\mu $, the
integration over $d\mu $ in Eq. (\ref{sigmaO1}) can be reduced to the
integration over $d\epsilon $. In the sums over $n$, only the term with $n=0$
remains because all $\omega _{n}$ with $n\neq 0$ are odd functions of $\mu $. 
If the Fermi surface has electron-like and hole-like pockets, the
conductivities are (compare with Ref. \cite{KL}) 
\begin{eqnarray}
\sigma _{O} &=&-\frac{|e|c}{B}\left[ N_{e}\int \left\langle \gamma
_{O}\right\rangle _{F,e}\frac{df_{0}}{d\epsilon }\,d\epsilon +N_{h}\int
\left\langle \gamma _{O}\right\rangle _{F,h}\frac{df_{0}}{d\epsilon }%
\,d\epsilon \right] ,  \nonumber \\
\sigma _{H} &=&-\frac{ec}{B}\left[ N_{e}\int \left\langle \gamma
_{H}\right\rangle _{F,e}\frac{df_{0}}{d\epsilon }\,d\epsilon -N_{h}\int
\left\langle \gamma _{H}\right\rangle _{F,h}\frac{df_{0}}{d\epsilon }%
\,d\epsilon \right] .  \nonumber
\end{eqnarray}

In the moderately clean case, the factors $\gamma _{O}$ and $\gamma _{H}$
are given by Eq. (\ref{gammas/mod}). The conductivities are 
\begin{eqnarray}
\sigma _{O} &\sim &\frac{Nec}{B}\,\frac{\Delta _{\infty }^{2}\tau }{E_{F}}%
\,\ln \left( \frac{T_{c}}{T}\right) \tanh \left( \frac{\Delta _{\infty }}{2T}%
\right) ;  \nonumber \\
~\frac{\sigma _{H}}{\sigma _{O}} &\sim &\frac{\Delta _{\infty }^{2}\tau }{%
E_{F}}\ln (T_{c}/T)  \label{sigma/mod}
\end{eqnarray}
where $N\sim N_{e}+N_{h}$. This is similar to the results for an $s$-wave
superconductor \cite{KopKr,KL,LO1}.

In the superclean limit $\Omega _{0}\tau \gg 1$ the factors $\gamma _{O}$
and $\gamma _{H}$ for low temperatures $T\ll T_{c}$ are determined by Eq. (%
\ref{gammas}). The general expression for the conductivities are rather
complicated. We consider two limits. The universal regime is reached when $%
\lambda \gg 1$, i.e., when either $\sqrt{H/H_{c2}}\ll 1/\left( \Omega
_{0}\tau \right) $ for $T/T_{c}\ll \sqrt{H/H_{c2}}$ or when $\omega _{c}\tau
\ll T/T_{c}$ for $T/T_{c}\gg \sqrt{H/H_{c2}}$. The region of the universal
regime is thus much larger than it was predicted in Ref. \cite{KopVol}. This
was pointed out by Makhlin \cite{Makhlin}. The condition $\lambda \gg 1$
automatically implies that $\omega _{c}\tau \ll 1$, therefore, both $d_{1}$
and $d_{2}$ are small. One has from Eq. (\ref{gammas}) $\gamma _{H}=1-2/\pi $%
, $\gamma _{O}=2/\pi $ which results in universal conductivities \cite
{KopVol}
\begin{eqnarray}
\sigma _{O} &=&\frac{|e|c}{B}(N_{e}+N_{h})\frac{2}{\pi };  \nonumber \\
~\sigma _{H} &=&\frac{ec}{B}(N_{e}-N_{h})\left( 1-\frac{2}{\pi }\right) .
\label{sigma/univ}
\end{eqnarray}

In the limit $\lambda \ll 1$ we have from Eq. (\ref{gammas/extreme}) 
\begin{equation}
\sigma _{O}=\frac{ec\left( N_{e}+N_{h}\right) }{B}~\frac{\Lambda }{\tau }%
;\;\sigma _{H}=\frac{\left( N_{e}-N_{h}\right) ec}{B}  \label{sigma/small}
\end{equation}
where $\Lambda =\left( N_{e}\Lambda _{e}+N_{h}\Lambda _{h}\right) \left(
N_{e}+N_{h}\right) ^{-1}$ and 
\begin{eqnarray}
\Lambda _{e,\,h} &=&\int_{0}^{\infty }\left\langle E_{{\rm col}%
}^{-1}\right\rangle _{Fe,\,Fh}\cosh ^{-2}\left( \frac{\epsilon }{2T}\right) 
\frac{d\epsilon }{2T}  \nonumber \\
&=&\left\langle \frac{1}{\sqrt{\Omega _{0}\omega _{c}}}+\frac{4T\ln 2}{\pi
\Delta _{\infty }\omega _{c}}\right\rangle _{Fe,F\,h}.
\end{eqnarray}
Here $\Omega _{0}$ is determined by Eq. (\ref{E/0}) with the logarithmic
factor $\ln [(T_{c}/T)\sqrt{H/H_{c2}}]$. The average is taken over the
particle-like and hole-like parts of the Fermi surface. Eq. (\ref
{sigma/small}) also agrees with \cite{Makhlin}. This regime is reached when
either $\sqrt{H/H_{c2}}\gg 1/\left( \Omega _{0}\tau \right) $ for $%
T/T_{c}\ll \sqrt{H/H_{c2}}$ or when $\omega _{c}\tau \gg T/T_{c}$ for $%
T/T_{c}\gg \sqrt{H/H_{c2}}$.

As the temperature approaches $T_{c}$, the contribution from fully
delocalized states with $\epsilon >\Delta _{\infty }$ becomes more and more
important. The delocalized states give the normal-state Ohmic and Hall
conductivities in the limit $T\rightarrow T_{c}$. In the limit $\omega
_{c}\tau \ll 1$, the flux-flow parts of $\sigma _{O}$ and $\sigma _{H}$
start from the universal values and then first decrease as $\Delta _{\infty
}/T_{c}\sim \sqrt{1-T/T_{c}}$ as long as $\Omega _{0}\tau $ remains large.
The ratio $\Delta _{\infty }/T_{c}$ measures the relative contribution from
localized states. With $T$ further approaching $T_{c}$, the moderately clean
regime is reached, and $\sigma _{O}$ becomes proportional to $(\Delta
_{\infty }/T_{c})\Omega _{0}\propto \left( 1-T/T_{c}\right) ^{3/2}$ while $%
\sigma _{H}\ $is proportional to $(\Delta _{\infty }/T_{c})\Omega
_{0}^{2}\propto \left( 1-T/T_{c}\right) ^{5/2}$ being small: $\sigma
_{H}/\sigma _{O}\sim \Omega _{0}\tau $ (compare with Ref. \cite{KL}). In the
limit $\omega _{c}\tau \gg 1$, the universal regime is not realized, and the
conductivity $\sigma _{O}$ starts from Eq. (\ref{sigma/small}) and saturates
at $\sigma _{O}\sim (Nec/B)(\omega _{c}\tau )^{-1}$ while $\sigma _{H}$
remains constant.

\subsection{Dispersion of conductivity}

\subsubsection{Low frequency: vortex mass}

If both the frequency and relaxation rate are low such that $\lambda
,\lambda ^{\prime }\ll 1$ the transverse response is $\left\langle \gamma
_{H}\right\rangle =1$. The transverse force is 
\begin{equation}
{\bf F}_{{\rm env,}\perp }=\pi \left( N_{e}-N_{h}\right) [\hat{{\bf z}}%
\times {\bf u}].  \label{F/trans}
\end{equation}
The longitudinal response,
\[
\left\langle \gamma _{O}\right\rangle =\frac{4}{\pi E_{{\rm col}}}\left(
-i\omega +\frac{1}{\tau }\right) ,
\]
gives the force 
\[
{\bf F}_{{\rm env,}\parallel }=-M\frac{\partial {\bf u}}{\partial t}-D{\bf u}
\]
where $D$ is the friction coefficient and $M$ plays the role of the vortex
``mass''. Indeed, the force balance takes the from of the Newton's law 
\begin{equation}
{\bf F}_{{\rm L}}+{\bf F}_{{\rm env,}\perp }-D{\bf u}=M\frac{\partial {\bf u}%
}{\partial t}.  \label{balance/mass}
\end{equation}
Here the vortex acceleration is a result of action of the Lorentz,
transverse, and friction forces. For a steady flux flow without a friction,
we would have from Eq. (\ref{balance/mass}) ${\bf j}_{s}=\left(
N_{e}-N_{h}\right) e{\bf u}$. In a system with the Galilean invariance,
where $N=N_{e}$ and $N_{h}=0$, the sum ${\bf F}_{{\rm L}}+{\bf F}_{{\rm env,}%
\perp }=\pi N[\left( {\bf v}_{s}-{\bf u}\right) \times \hat{{\bf z}}]$ is
the Magnus force. The friction coefficient and the mass per unit length are 
\[
D=\pi \left( N_{e}+N_{h}\right) \Lambda /\tau ;\;M=\pi \left(
N_{e}+N_{h}\right) \Lambda .
\]
The effective mass of a $d$-wave vortex 
\[
M\sim Nm\xi ^{2}\left( \sqrt{\frac{H_{c2}}{H}}+\frac{T}{T_{c}}\frac{H_{c2}}{H%
}\right) 
\]
is much larger than the mass of a conventional vortex \cite{Kopnin/mass} $%
M\sim Nm\xi ^{2}$ obtained in the same way. For temperatures $T/T_{c}\ll 
\sqrt{H/H_{c2}}$ , it is the mass \cite{Volovik/mass} of electrons inside
the ``vortex core'' with the dimensions $\xi $ by $\xi \sqrt{H_{c2}/H}$. The
mass increases with temperature. Note that this mass is very large compared
with what is usually obtained by calculating the corrections to the kinetic
energy of a moving vortex proportional to the second power of its velocity
including the electromagnetic energy. For example, the recent result of Ref. 
\cite{Gaitonde} is roughly by the factor of $(T_{c}/E_{F})^{4}$ smaller! We
have to stress that this mass appears as a response to a slow acceleration.
With an increase in the characteristic frequency of the vortex motion, the
response transforms into a highly dissipative resonant behavior where a mass
is not an appropriate quantity.

\subsubsection{High frequency: resonances.}

Due to the poles in $\left\langle \gamma _{O}\right\rangle $, the
dissipative component of the response has a real part even for $\tau \to
\infty $. For $T\ll \Delta _{\infty }\sqrt{\omega _{c}/\Omega _{0}}$, the
absorption is concentrated near $\omega _{k}(p_{z})=(2k+1)E_{0}(p_{z})$ with
the line width $\delta \omega _{k}/\omega _{k}\sim (T/\Delta _{\infty })%
\sqrt{\Omega _{0}/\omega _{c}}\ll 1$: 
\[
{\rm Re}\,\sigma _{O}=\frac{Nec}{B}\sum_{k}\left\langle \frac{4E_{0}}{\pi }%
\delta (\omega -\omega _{k})\right\rangle _{F}.
\]
Averaging over the Fermi surface, i.e., integration over $dp_{z}$, gives
rise to the van Hove singularities in conductivity at frequencies where $%
\partial E_{0}/\partial p_{z}=0$, i.e., where $\omega =(2k+1)E_{0}(p_{z}=0)$.

For $T\gg \Delta _{\infty }\sqrt{\omega _{c}/\Omega _{0}}$, we have 
\begin{equation}
{\rm Re}\,\sigma _{O}=\frac{Nec}{B}\left\langle \frac{\Delta _{\infty
}\omega _{c}}{T\omega }\sum_{k\geq 0}\cosh ^{-2}\left[ \frac{\pi \Delta
_{\infty }\omega _{c}}{4T\omega }(1+2k)\right] \right\rangle _{F}.
\label{sigma/res}
\end{equation}
The conductivity is exponentially small for low frequency $\omega \ll
E_{g}=\pi \Delta _{\infty }\omega _{c}/4T$. It starts to rise at the
absorption edge $\omega \sim E_{g}$. The average in the r.h.s. of Eq. (\ref
{sigma/res}) is equal to $2/\pi $ for large frequencies $\omega \gg E_{g}$,
when transitions between many resonance modes are excited. This is
equivalent to the limit $\lambda \gg 1$ for a steady flow.

The absorption occurs at the collective modes of electrons involved into the
vortex motion. For temperatures $T\ll T_{c}\sqrt{H/H_{c2}}$, these modes are
transitions between the true quantum mechanical energy levels determined by
Eq. (\ref{BS/quant}). For higher temperatures, delocalized quasiparticles
are involved into the vortex motion, and the resonances take place at
collective modes with energies of the order of $E_{g}$.

\section{Summary}

For a steady vortex motion, one can identify three regimes in the order of
increasing $\tau $: (i) The moderately clean regime with $\Omega _{0}\tau
\ll 1$. In this case, the main response of vortices to the external current
is dissipative with only a small Hall angle proportional to $\Omega _{0}\tau 
$. The conductivities behave similarly to the $s$-wave case, see Eq. (\ref
{sigma/mod}). Next comes the superclean limit $\Omega _{0}\tau \gg 1$.
However, in $d$-wave superconductors, it further separates into two
sub-regimes. The relevant parameter is the energy of collective modes
excited by the moving vortex. The vortex dynamics in a $d$-wave
superconductor depends crucially on the behavior of excitations near the gap
nodes. Due to the presence of gap nodes, only the excitations with very low
energies $\epsilon \ll \Delta _{\infty }\sqrt{H/H_{c2}}$ are localized in a
vortex core. The excitations with higher energies, coming into play for
temperatures above $T_{c}\sqrt{H/H_{c2}}$, are actually the collective modes
where both classically localized and delocalized particles participate. The
interplay between the energy of these collective modes $E_{{\rm col}}$ and
the relaxation rate or the frequency of vortex oscillations determines which
of the two sub-limits is realized: (ii) Intermediate or ``universal'' regime
with $E_{{\rm col}}\tau \ll 1$ but $\Omega _{0}\tau \gg 1$. Here both
dissipative and Hall components of conductivity tensor are universal. They
are independent of the relaxation time and are only determined by the
magnetic field and by the number of electrons and holes under the Fermi
surface in the normal state, Eq. (\ref{sigma/univ}). (iii) Extremely clean
limit $E_{{\rm col}}\tau \gg 1$ where the dissipative part of the vortex
response is small, and the transverse Hall component of conductivity
dominates.

Provided the superclean condition $\Omega _{0}\tau \gg 1$ is satisfied, the
universal regime can be realized for temperatures $T/T_{c}\ll \sqrt{H/H_{c2}}
$ under the condition $E_{0}\tau \ll 1$, i.e., $\Omega _{0}\tau \ll \sqrt{%
H_{c2}/H}$. This condition for the universal regime was predicted in Ref. 
\cite{KopVol}. However, as was noticed in Ref. \cite{Makhlin}, the universal
regime is not restricted to this temperature range. It can also be realized
for higher temperatures if $\omega _{c}\tau \ll T/T_{c}$. For these
temperatures, the dissipative conductivity vanishes only in the extreme
clean limit $\omega _{c}\tau \gg T/T_{c}$.

The temperature and magnetic-field dependences of conductivities in the
extremely clean limit are quite unusual. Extremely clean limit can be
reached by increasing the magnetic field above $H/H_{c2}\sim (\Omega
_{0}\tau )^{-2}$ for $T/T_{c}\ll (\Omega _{0}\tau )^{-1}$ or above $%
H/H_{c2}\sim (\Omega _{0}\tau )^{-1}(T/T_{c})$ for temperatures $(\Omega
_{0}\tau )^{-1}\ll T/T_{c}<1$. At the crossover from universal to extremely
clean regime, the Hall conductivity grows by the factor $\pi /(\pi
-2)\approx 2.75$ while the Ohmic conductivity decreases, Eq. (\ref
{sigma/small}). For temperatures $T/T_{c}\ll \sqrt{H/H_{c2}}$, its field
dependence is \cite{KopVol} $\sigma _{O}\propto H^{-3/2}$; it transforms
into \cite{Makhlin} $\sigma _{O}\propto H^{-2}$ for $T/T_{c}\gg \sqrt{%
H/H_{c2}}$. The Hall angle approaches $\pi /2$.

At temperatures $T\sim T_{c}$ but not specifically close to $T_{c}$, both
Ohmic and Hall conductivities are of the order of $Nec/B$ as long as $\omega
_{c}\tau \ll 1$. They, of course, do not have the universal values any more
because all angles contribute similarly to the conductivities. However, the
main conclusion is qualitatively the same: dissipation only disappears when $%
\omega _{c}\tau \gg 1$. This is the major difference between $s$-wave and $d$%
-wave superconductors: in former, dissipation vanishes already for $\Omega
_{0}\tau \gg 1$. For $T\to T_{c}$, the superclean condition $\Delta _{\infty
}^{2}\tau /E_{F}\gg 1$ fails, and one approaches the moderately clean limit
where dissipation dominates.

Having excited an oscillatory motion of vortices by a microwave irradiation
at temperatures $T/T_{c}\ll \sqrt{H/H_{c2}}$ with the frequency $\omega \sim
(\Delta _{\infty }^{2}/E_{F})\sqrt{H/H_{c2}}$, one can observe absorption
resonances at the vortex-core states, if the condition $\omega \tau \gg 1$
is satisfied. The required frequency is higher than the cyclotron frequency
and has a $\sqrt{H}$ dependence on the magnetic field. For higher
temperatures $T/T_{c}\gg \sqrt{H/H_{c2}}$, one observes an absorption edge
at a frequency of the order of $(T_{c}/T)\omega _{c}$ where the absorption
increases sharply with the increasing frequency due to resonances at the
vortex collective modes.

Finally, we find that the response of a vortex to a slow acceleration is
characterized by a ``mass'' term. The mass per unit length, by the order of
magnitude, is equal to the mass of electrons inside the vortex core with
dimensions $\xi $ by $\xi \sqrt{H_{c2}/H}$ and increases with temperature,
i.e., it is enormously larger than the electromagnetic contribution or any
other correction to the kinetic energy of a vortex proportional to the
second power of its velocity.

\acknowledgements

I am grateful to Yu. G. Makhlin and G. E. Volovik for illuminating
discussions. This work was supported the Swiss National Foundation
cooperation grant 7SUP J048531 and by the INTAS grant 96-0610. The support
by the Russian Foundation for Fundamental Research grant No. 96-02-16072 and
by the program ``Statistical Physics'' of the Ministry of Science of Russia
is also acknowledged.

\appendix

\section{Quasiclassical spectrum in $d$-wave superconductors}

The quasiclassical approximation, a quasiparticle passing near a vortex has
a definite trajectory which is actually a straight line characterized by an
impact parameter $b$. This argumentation holds for any symmetry of the order
parameter and can be applied to a $d$ -wave superconductor, as well. We
solve the quasiclassical Eilenberger equations for the Green functions $g$, $%
f$, and $f^{\dagger }$ using the same scheme as in Ref. \cite{KrPe}. In the
frame with the impact parameter $b$ and the distance along the trajectory $s$
as the coordinates introduced in Section \ref{excitations}, the Eilenberger
equations are 
\begin{eqnarray}
-iv_{\perp }\frac{\partial f}{\partial s}-2\tilde{\epsilon}f+2\Delta _{{\bf p%
}}g &=&0~;  \label{frp1} \\
iv_{F}\frac{\partial f^{\dagger }}{\partial s}-2\tilde{\epsilon}f^{\dagger
}+2\Delta _{{\bf p}}^{*}g &=&0~.  \label{+frp1}
\end{eqnarray}
and $g^{2}-f\,f^{\dagger }=1$. Here the functions $g$, $f$, and $f^{\dagger
} $ are either retarded or advanced Green's functions. The energy $\tilde{%
\epsilon}=\epsilon +(2e/c)v_{\perp }A_{s}$. For an extreme type II
superconductor with the Ginzburg--Landau parameter $\kappa \gg 1$, the
magnetic field is nearly homogeneous, $H=n_{L}\phi _{0}$ with $A_{\phi
}=H\rho /2$. Since the vector potential component $A_{s}=-(b/\rho )A_{\phi }$
we have $\tilde{\epsilon}=\epsilon -eHbv_{\perp }/2c$.

The order parameter of a $d$-wave superconductor has the form of Eq. (\ref
{order/param}) We introduce 
\begin{equation}
\eta =-\int_{0}^{s}{\frac{{\partial \phi }}{{\partial s}}}\,ds=\frac{\pi }{2}%
+\alpha -\phi (s).
\end{equation}
so that $\cos \eta =b/\rho $ and $\sin \eta =s/\rho $. We put 
\begin{equation}
~f=f_{0}\exp (i\phi +i\eta );\;f^{\dagger }=f_{0}^{\dagger }\exp (-i\phi
-i\eta ),
\end{equation}
and 
\begin{equation}
f_{0}=-\theta (s)+i\,\zeta (s);~f_{0}^{\dagger }=\theta (s)+i\,\zeta (s);
\label{thetazeta}
\end{equation}
The functions $\theta $ and $\zeta $ satisfy the equations 
\begin{eqnarray}
v_{\perp }{\frac{{\partial \zeta }}{{\partial s}}}+2\tilde{\epsilon}\theta
-2i\Delta _{0{\bf p}}g\sin \eta &=&0;  \label{zetad} \\
v_{\perp }{\frac{{\partial \theta }}{{\partial s}}}-2\tilde{\epsilon}\zeta
-2i\Delta _{0{\bf p}}g\cos \eta &=&0.  \label{thetad}
\end{eqnarray}
where $g^{2}=1-\zeta ^{2}-\theta ^{2}$.

For energies close to the eigen energy, i.e., near the poles, the Green
functions $f$ and $g$ are large. The function $g$ should be proportional to $%
\zeta $ which has the required symmetry with respect to $s$. Therefore, we
assume that 
\begin{equation}
\zeta ^{2}\gg 1-\theta ^{2}  \label{applcond}
\end{equation}
so that $g=i\sqrt{\zeta ^{2}+\theta ^{2}-1}\approx i\zeta $. The sign is
chosen to ensure a decay of $g$ at large distances from the vortex. The
function $\zeta $, being the amplitude of the residue of $g$ for the pole,
decreases and should completely vanish at large distances.. It is reasonable
to assume that, at large distances, $\zeta \to 0$ and $\theta ^{2}\to 1$. At
large distances from the vortex axis, the full Green function 
\begin{equation}
f^{R}=-i\frac{\Delta _{{\bf p}}}{\sqrt{|\Delta _{{\bf p}}|^{2}-\tilde{%
\epsilon}^{2}}}.  \label{f/large/dist}
\end{equation}
In combination with Eq. (\ref{thetazeta}), it gives the required sign of $%
\theta $. Finally, the boundary condition for the Eilenberger equations are 
\begin{equation}
\zeta \to 0;~\theta \to \pm {\rm sign}\,[\sin (2\alpha )]~{\rm for}~s\to \pm
\infty .  \label{bound/cond}
\end{equation}

The solution is 
\begin{eqnarray}
\zeta ^{R(A)} &=&\frac{{\rm sign}\,[\sin (2\alpha )]\,v_{\perp }e^{-K}}{%
\int_{-\infty }^{\infty }\left[ \tilde{\epsilon}-\Delta _{0}|\sin (2\alpha
)|\cos \eta \pm i\delta \right] e^{-K}\,ds}  \nonumber \\
\theta ^{R(A)} &=&2{\rm sign}\,[\sin (2\alpha )]  \nonumber \\
&\times &\left( \frac{\int_{-\infty }^{s}\left[ \tilde{\epsilon}-\Delta
_{0}|\sin (2\alpha )|\cos \eta \right] e^{-K}\,ds^{\prime }}{\int_{-\infty
}^{\infty }\left[ \tilde{\epsilon}-\Delta _{0}|\sin (2\alpha )|\cos \eta \pm
i\delta \right] e^{-K}\,ds}-\frac{1}{2}\right)  \nonumber
\end{eqnarray}
where 
\begin{equation}
K=\frac{2}{v_{\perp }}\int_{s_{0}}^{s}\Delta _{0}|\sin (2\alpha )|\sin \eta
\,ds^{\prime }.
\end{equation}
As a result, 
\begin{equation}
g^{R(A)}=\frac{iv_{\perp }e^{-K}}{C\left[ \epsilon -\epsilon _{0}\pm i\delta
\right] }
\end{equation}
where 
\begin{eqnarray}
\epsilon _{0}(\alpha ,b) &=&C^{-1}|\sin (2\alpha )|\int_{-\infty }^{\infty
}\Delta _{0}\cos \eta e^{-K}\,ds  \nonumber \\
&+&eHbv_{\perp }/2c  \label{Eb}
\end{eqnarray}
with 
\begin{equation}
C=\int_{-\infty }^{\infty }e^{-K}\,ds.  \label{C}
\end{equation}
Eq. (\ref{Eb}) was obtained in \cite{deG,KrPe} for an $s$-wave
superconductor.

For small impact parameters, $b\ll \xi $, one has $\mid s\mid =\rho $.
Therefore, 
\begin{equation}
\epsilon _{0}(b)=b\left( C^{-1}|\sin (2\alpha )|\int_{0}^{\infty }\Delta
_{0}e^{-K(\rho )}\,\frac{d\rho }{\rho }+p_{\perp }\omega _{c}/2\right) .
\label{Eb2}
\end{equation}
where $\omega _{c}=|e|H/m_{\perp }c$ is the cyclotron frequency with $%
m_{\perp }=p_{\perp }/v_{\perp }$. Modulus of charge appears due to the
choice of the $z$ axis. Eq. (\ref{Eb2}) was obtained in \cite{Hansen}. Due
to the Kramer and Pesch effect \cite{KrPe}, the core size shrinks at low
temperatures to $\xi _{1}\sim (T/T_{c})\xi $. The spectrum takes the form of
Eq. (\ref{E/small/b}) with $L=\ln \,[1/|\sin (2\alpha )|]$.

The contribution from the magnetic field increases if the number of vortices
increases together with the magnetic field. However, the order-parameter
phase gradient $\nabla \chi $ at the position of the vortex core under
consideration remains equal to that produced by a single vortex $\nabla
_{\phi }\chi =1/\rho $ as long as $\rho \ll \rho _{\max }$ where $\rho
_{\max }$ is the intervortex distance since contributions from other
vortices cancel each other.

Near the gap nodes, where $\sin (2\alpha )\ll 1$, Eq. (\ref{Eb}) for the
anomalous branch can be simplified. Significant values of $K$ are determined
by $\rho \gg \xi $. Therefore, one can put $\Delta _{0}=\Delta _{\infty }$: 
\begin{eqnarray*}
K(s) &=&\frac{2\Delta _{\infty }|\sin (2\alpha )|}{v_{\perp }}\int_{0}^{s}%
\frac{s^{\prime }\,ds^{\prime }}{\rho ^{\prime }} \\
&=&\frac{2\Delta _{\infty }|\sin (2\alpha )|}{v_{\perp }}(\rho -b)
\end{eqnarray*}
The normalization constant is 
\begin{equation}
C=2e^{\gamma }\int_{b}^{\infty }e^{-\gamma \rho /b}\frac{\rho \,d\rho }{%
\sqrt{\rho ^{2}-b^{2}}}=be^{\gamma }K_{1}(\gamma )
\end{equation}
where $K_{1}$ is the Bessel function of an imaginary argument with $\gamma
=2\Delta _{\infty }b|\sin (2\alpha )|/v_{\perp }$. The energy becomes 
\begin{equation}
\epsilon _{0}(b)=\Delta _{0}|\sin 2\alpha |\frac{K_{0}(\gamma )}{%
K_{1}(\gamma )}+bp_{\perp }\omega _{c}/2.  \label{E/nodes}
\end{equation}
As a result, for $\xi \ll b\ll \xi /|\sin 2\alpha |$, we get Eq. (\ref
{E/small/b}) with $L=\ln [\xi v_{\perp }/bv_{F}|\sin (2\alpha )|]$. The
low-lying states with energies much below the gap at infinity, $\epsilon \ll
\Delta _{0}|\sin 2\alpha |$ correspond to $b\ll \xi /|\sin 2\alpha |$. For
large impact parameters, $b\gg \xi /|\sin 2\alpha |$, we can use the
asymptotics of the Bessel functions for large $\gamma $ and arrive at Eq. (%
\ref{spec/nodes3}).

Eq. (\ref{E/nodes}) holds when Eq. (\ref{applcond}) is fulfilled which is
equivalent to 
\begin{equation}
\int_{0}^{s}(\tilde{\epsilon}-\Delta _{0}\cos \eta )e^{-K}\,ds\ll v_{\perp }
\label{condition}
\end{equation}
for such $s$ that $K\sim 1$. Eq. (\ref{condition}) is satisfied either for $%
\gamma \ll 1$ or for $\gamma \gg 1$. In the latter case, the exponent is
localized at distances $s\sim \sqrt{b\xi /|\sin (2\alpha )|}\ll b$. Using
Eq. (\ref{spec/nodes3}) we find that the integral in Eq. (\ref{condition})
is of the order of $v_{\perp }s/b\sim v_{\perp }\sqrt{\xi /b|\sin (2\alpha )|%
}\ll v_{\perp }$. We see that for $b\gg \xi /|\sin (2\alpha )|$, a particle
is localized at $\cos \eta =1$. i.e., on a line passing through the vortex
perpendicular to the momentum direction.

The energy levels for $n\neq 0$ can be found using the full quasiclassical
scheme. It works either for $b\gg \xi $ or for $n\gg 1$. We look for a
solution for Bogoliubov wave functions in the form $e^{i\int p\,d\rho }$ .
The quasiclassical momentum is 
\begin{equation}
p=q_{\rho }\pm \frac{m}{q_{\rho }}\sqrt{\left( \tilde{\epsilon}+\frac{%
bv_{\bot }}{2\rho ^{2}}\right) ^{2}-\Delta _{0}^{2}\sin ^{2}(2\alpha )}
\label{p/quasi}
\end{equation}
where $q_{\rho }=p_{\perp }\sqrt{1-b^{2}/\rho ^{2}}$. The integral along the
closed quasiparticle trajectory is 
\begin{eqnarray}
\oint p\,d\rho &=&4\int_{b}^{\rho _{0}}\frac{m}{q_{\rho }}\sqrt{\left( 
\tilde{\epsilon}+\frac{bv_{\perp }}{2\rho ^{2}}\right) ^{2}-\Delta
_{0}^{2}\sin ^{2}(2\alpha )}\,d\rho  \nonumber \\
&=&2\pi \left( n+\frac{1}{2}\right)  \label{high/n/quant}
\end{eqnarray}

Consider positive and large $b\gg \xi $ and positive energies $\tilde{%
\epsilon}\to \Delta _{\infty }|\sin (2\alpha )|$. At large distances, $%
1-\Delta _{0}/\Delta _{\infty }\sim \xi ^{2}/\rho ^{2}$. For $b\gg \xi $ ,
the term with $b$ in Eq. (\ref{high/n/quant}) dominates over the
large-distance correction to $\Delta _{\infty }$. The zero of the square
root in $p$ determines the turning point 
\[
\rho _{0}=\sqrt{\frac{bv_{\bot }}{2(\Delta _{\infty }|\sin (2\alpha )|-%
\tilde{\epsilon})}}. 
\]
For $\rho _{0}\gg b$ , the integral in Eq. (\ref{high/n/quant}) is
logarithmic 
\begin{equation}
\oint p\,d\rho =4\sqrt{\frac{b\Delta _{\infty }|\sin (2\alpha )|}{v_{\bot }}}%
\ln \frac{\rho _{0}}{b}.  \label{quant1}
\end{equation}
Large $\rho _{0}/b$ correspond to $x_{0}\gg 1$ where 
\[
x_{0}=\pi \left( n+\frac{1}{2}\right) \sqrt{\frac{v_{\bot }}{b\Delta
_{\infty }|\sin (2\alpha )|}}. 
\]
In this limit, 
\begin{equation}
\tilde{\epsilon}_{n}=\Delta _{\infty }|\sin (2\alpha )|-\frac{v_{\perp }}{2b}%
\,e^{-x_{0}}.  \label{En}
\end{equation}
Thus, for large $n$ and small enough $b\ll \pi ^{2}n^{2}\xi /|\sin (2\alpha
)|$, the energy is exponentially close to the gap $\Delta _{\infty }|\sin
(2\alpha )|$. The derivative with respect to the impact parameter is
negative 
\[
\frac{\partial \tilde{\epsilon}_{n}}{\partial b}=-\frac{v_{\perp }x_{0}}{%
4b^{2}}\,e^{-x_{0}}. 
\]
The separation between the gap and $\tilde{\epsilon}$ increases with $b$
until $x_{0}\sim 1$. The derivative becomes positive for larger impact
parameters when $\rho _{0}\rightarrow b$, i.e., when $x_{0}\ll 1$. The
energy is 
\[
\tilde{\epsilon}_{n}=\Delta _{\infty }|\sin (2\alpha )|-\frac{v_{\perp }}{2b}%
. 
\]
The correction to $\Delta $ is two times larger than that obtained for $%
\epsilon _{0}(b)$ from the more exact Eq. (\ref{spec/nodes3}). This is
because the full quasiclassical approach does not work well near the turning
point. The minimum of energy is reached at $x_{0}\sim 1$ and is equal to 
\[
\tilde{\epsilon}_{n}\approx \Delta _{\infty }|\sin (2\alpha )|\left( 1-\frac{%
1}{2\pi ^{2}n^{2}}\right) 
\]
The number of branches with the energy between $\tilde{\epsilon}$ and $%
\Delta _{\infty }$ is thus given by Eq. (\ref{Number}).

\end{document}